\documentclass[aps,showpacs,showkeys,twocolumn,prb,amsmath,amssymb,superscriptaddress]{revtex4-1}
\usepackage[english]{babel}
\usepackage{color}
\usepackage{graphicx}
\usepackage{graphics}
\begin{document}
%
%----------------------------head----------------------------------
%
\title{The phase transition in the anisotropic Heisenberg model with
long range dipolar interactions}
\author{L.A.S. M\'ol}
  \email{lucasmol@fisica.ufmg.br}
  \author{B.V. Costa}
 \email{bvc@fisica.ufmg.br}
    \affiliation{ Departamento de F\'isica, Laborat\'orio de
Simula\c{c}\~ao, ICEX, UFMG, 30123-970, Belo Horizonte, MG, Brazil }
\date{\today}
\begin{abstract}
In this work we have used extensive Monte Carlo calculations to study the planar to paramagnetic phase transition in the two-dimensional anisotropic Heisenberg model with dipolar interactions (AHd) considering the true long-range character of the dipolar interactions by means of the Ewald summation. Our results are consistent with an order-disorder phase transition with unusual critical exponents in agreement with our previous results for the Planar Rotator model with dipolar interactions. Nevertheless, our results disagrees with the Renormalization Group results of Maier and Schwabl [PRB, {\bf 70}, 134430 (2004)] and the results of Rapini et. al. [PRB, {\bf 75}, 014425 (2007)], where the AHd was studied using a cut-off in the evaluation of the dipolar interactions. We argue that besides the long-range character of dipolar interactions their anisotropic character may have a deeper effect in the system than previously believed. Besides, our results shows that the use of a cut-off radius in the evaluation of dipolar interactions must be avoided when analyzing the critical behavior of magnetic systems, since it may lead to erroneous results.
\end{abstract}
\pacs{75.10.Hk, 75.30.Kz, 75.40.Mg}
\keywords{Classical spin models; Monte Carlo simulations; Phase transitions; Dipolar interactions; Long range interactions}
\maketitle
%
%----------------------------Section----------------------------------
%
\section{\label{introducao}Introduction}

A wide class of interesting phenomena is observed in quasi-two dimensional systems like thin films, surfaces, superconductors and easy plane magnets. For example, since the work of Mermin and Wagner \cite{mermin-wagner} it is known that a continuous symmetry cannot be spontaneously broken at finite temperature in systems with sufficiently short-range interactions in dimensions $d \leq 2$. Although an order-disorder transition  is forbidden in two dimensions a non-usual phase transition is still possible as pointed by Berezinskii \cite{berezinskii} and Kosterlitz and Thouless \cite{kosterlitz} (BKT). A prototype model undergoing a BKT transition is the 2d easy-plane anisotropic Heisenberg (2dAH) model described by the following Hamiltonian
\begin{equation}\label{eq-1}
    H_{0}  = -J\sum_{i,j} \vec{S}_i \cdot \vec{S}_j - A\sum_i \left(  S^z_i \right)^2 ,
\end{equation}
\noindent
with $A < 0$. If $A> 0$ the hamiltonian has an easy axis symmetry being in the Ising class of universality. For $A=0$ the model turns to the isotropic Heisenberg model which is known to have no phase transition at all. The addition of long range interactions having a power-law fall off in two spatial dimensions may lead to significant changes in the character of the phase transition. As discussed by Fisher et al.~\cite{fisher-1} long range attractive interactions should lead to modification in the values of the critical exponents from those of the corresponding models with short range interactions. Such interaction potentials can induce critical behavior in dimensions smaller than or equal two. 

\begin{figure}
 \includegraphics[scale=0.3]{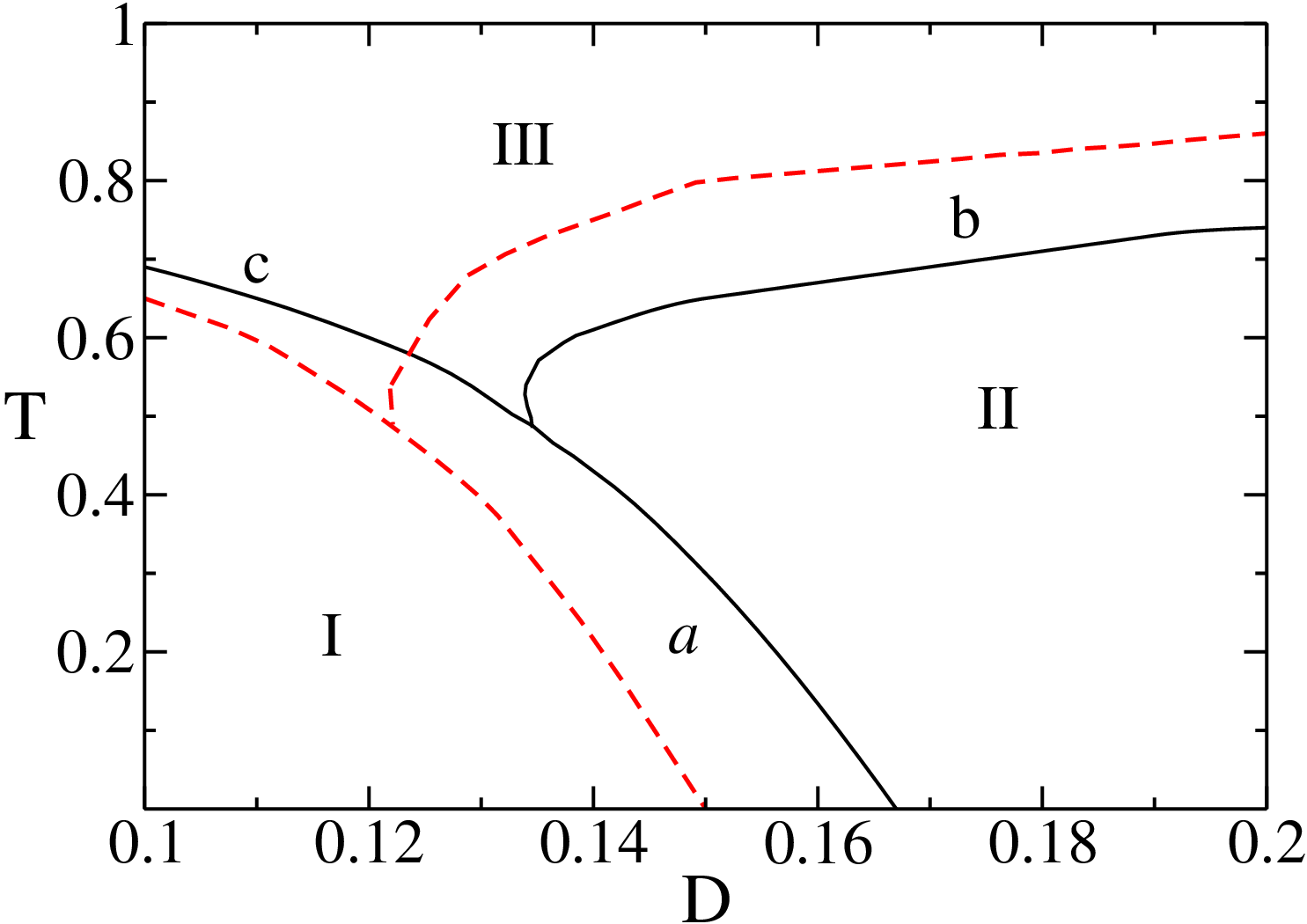} \\
 \caption{\label{fig-1} (Color online) Phase diagram of the anisotropic Heisenberg model with dipolar
interactions (AHd) for fixed $A/J=1$ in the $(D/J,T)$ space. The black solid line
represents the transition lines as obtained using a cut-off in the dipolar interactions
\cite{marcella} and the solid black line are the results obtained when full long-range interactions
are considered by means of the Ewald summation (this work). The phase I is an Ising-like phase characterized
by an ordered out-of-plane alignment of spins (that may present stripe-like configurations for full
long-rang interactions). Phase II
is an ordered planar ferromagnetic state and phase III is a paramagnetic one. The border
line between phase I and phase II ($a$) is believed to be of first order and
from regions I and II to III ($b$ and $c$) to be both of second order.}
\end{figure}

In a real magnet the long range dipolar interactions between magnetic moments are always present. Then, a better model designed to describe real magnets is 
\begin{equation}\label{eq-2}
    H_{d}  = H_{0} + D\sum_{(i,j)}\left[ \frac{\vec{S}_i \cdot \vec{S}_j}{r^3_{i,j}} - \frac{3\left( \vec{S}_i \cdot \vec{r}_{i,j}\right)\left( \vec{S}_j \cdot \vec{r}_{i,j}\right)}{r^5_{i,j}}   \right],
\end{equation}
\noindent
which is named anisotropic Heisenberg model with dipolar interactions (AHd). Despite the fundamental relevance of the theoretical problem the technological interest in low dimensional systems with long range interactions makes the study of such models of paramount importance. The inclusion of dipolar interactions induces the appearance of an easy-plane anisotropy in quasi-two dimensional systems in such a way that for $A>0$, i.e., for an easy-axis site anisotropy, the competition between them leads to interesting phenomena. In earlier studies several authors~\cite{debell,bruno,taylor,whitehead,carubelli,santamaria,macisaac} have claimed that the model for ultrathin magnetic films defined by Eq. \ref{eq-2} with $A>0$ presents three phases. Referring to Fig. \ref{fig-1} it is believed that the line labeled $a$ is of first order. The lines $b$ and $c$ are of second order. Those results were obtained by introducing a cutoff in the long-range interaction of the Hamiltonian, $H_{d}$.

 An attempt to determine the true character of the planar-to-paramagnetic phase (line $b$ in figure~\ref{fig-1}) was done by Maier and Schwabl~\cite{maier}. In their work, the authors used renormalization group technique to study the model with dipolar interactions (Eq. \ref{eq-2}). They discussed the existence of a new universality class with characteristics of $BKT$ and order-disorder transitions as well. They argued that the dipolar $XY$ model exhibit long-range order at low temperature (see also Refs.~\onlinecite{maleev,bruno}), but the correlation length diverges exponentially as the critical temperature is approached. The specific heat does not present any divergence as in a $BKT$ transition. The susceptibility was expected to diverge as $\chi \propto \xi^{\tilde{\gamma}}$ where $\tilde{\gamma} = \gamma/\nu = 1$ is the critical susceptibility exponent and $\chi$ is the correlation length. The magnetization approaches zero as $M \propto \xi^{-\tilde{\beta}}$, where $\tilde{\beta} = \beta/\nu = 1/2$, and the correlation function exponent was found to be $\tilde{\eta}=\eta/\nu = 1$ while in the $BKT$ picture is found $\tilde{\gamma} = 7/4$ and $\tilde{\eta} = 1/4$. They stated that~\cite{maier} `` The nature and flow diagram of the ferromagnetic transition are strikingly similar to the Kosterlitz-Thouless transition. But while in the Kosterlitz-Thouless transition the exponential behavior is a consequence of the topological excitations, the predicted phenomena in the dipolar $XY$ model are solely due to spin-wave excitations ''. The Maier and Schwabl result, in some sense, corroborate the findings of an earlier work of Patrascioiu and Seiler~\cite{patrascioiu}.  Their results~\cite{patrascioiu} `` ... lead to an interpretation of the KosterlitzThouless transition, different from the standard one, of dipole dissociation. ''.

Using numerical Monte Carlo (MC) calculations Rapini et al.~\cite{marcella} have found that the line labeled $a$ is of first order and the line $c$ is of second order in agreement with Ref.~\onlinecite{santamaria}. However, the $b$ line was found to be of the $BKT$ type. M\'ol and Costa have used extensive Monte Carlo simulations and finite-size scaling theory to study the planar to paramagnetic transition (Line $b$ in figure \ref{fig-1}) in two versions of this model: The first was a bilayer version using a cutoff in the dipole interaction~\cite{mol2p}, the second was the dipole planar rotator model, where the spins have a $O(2)$ symmetry~\cite{molxy}. In the last case the dipole interaction was considered without a cutoff by using the Ewald summation technique. In both cases the $b$ line was not found to be of second order neither of $BKT$ type. In particular they found that both transitions might belong to a peculiar universality class. The results indicated that the transition is characterized by a non-divergent specific heat and by the exponents $\beta =0.18(5)$, $\gamma =2.1(2)$ and $\nu =1.22(9)$ in the bilayer case~\cite{mol2p} and $\beta =0.2065(4)$, $\gamma =2.218(5)$, and $\nu =1.277(2)$ in the planar rotator model~\cite{molxy}. These results are far different of those predicted by the $BKT$ theory but closer to the Maier and Schwabl's results~\cite{maier}. As a step further to shed some light over this question we have done a very careful MC study of the anisotropic Heisenberg model with dipolar interactions (AHd). The technical details and the results we have obtained are presented in the following.

\section{\label{MC}Numerical Details}

The Monte Carlo scheme we used was a plain single site canonical Metropolis algorithm since conventional cluster algorithms cannot be used due to the long-range anisotropic character of the dipolar interactions. The Metropolis algorithm is sufficiently well known to deserve any further presentation. We define a Monte Carlo step ($MCS$) as consisting of an attempt to assign a new random direction to all spins in the lattice. To equilibrate the system we used $100 \times L^2 MCS$ which was found to be sufficient to reach equilibrium even in the vicinity of the phase transition. We produced histograms for each lattice size in the interval $20 \leq L \leq 120$ and they were built at/close to the estimated critical temperatures obtained in preliminary  simulations. To construct the histograms at least $2 \times 10^7$ configurations were obtained using 3 distinct runs. These histograms are summed so that we obtain a new histogram that allow us to explore a wider range of temperature (an example of the use of histograms can be found in Ref. \onlinecite{mol2p}). Periodic boundary conditions are assumed in the $x$ and $y$ directions. To take into account the long range character of the dipolar interaction we use the Ewald summation to calculate the energy of the system\cite{weis,wang}.

All simulations where done using a square lattice, $A/J=1$ and $D=0.3J$. Energy was measured in units of $J$ and temperature in units of $J/k_B$, where $k_B$ is the Boltzmann constant. Our choice of $D=0.3J$ was to guarantee that the planar behavior of the system was not much affected by the frustration existent  near the multicritical point where the three lines shown in Fig. \ref{fig-1} come together. We have devoted our efforts to determine a number of thermodynamic quantities, namely the specific heat, magnetization, susceptibility, fourth order Binder's cumulant and moments of magnetization as described elsewhere~\cite{mol2p,molxy}.

\section{\label{Results}Results}

Concerning the systems' magnetization no significant
size dependence is observed in low temperatures, unlike the results shown in
Ref.~\onlinecite{marcella} where a cut-off radius were used in the evaluation
of dipolar interactions. This may be an evidence that as the full long-range 
character of dipolar interactions are taken into account long-range order
develops, as expected by the results of Maleev\cite{maleev}.

\begin{figure}
%\vspace{0.5cm}
\begin{center}
\includegraphics[scale=0.25]{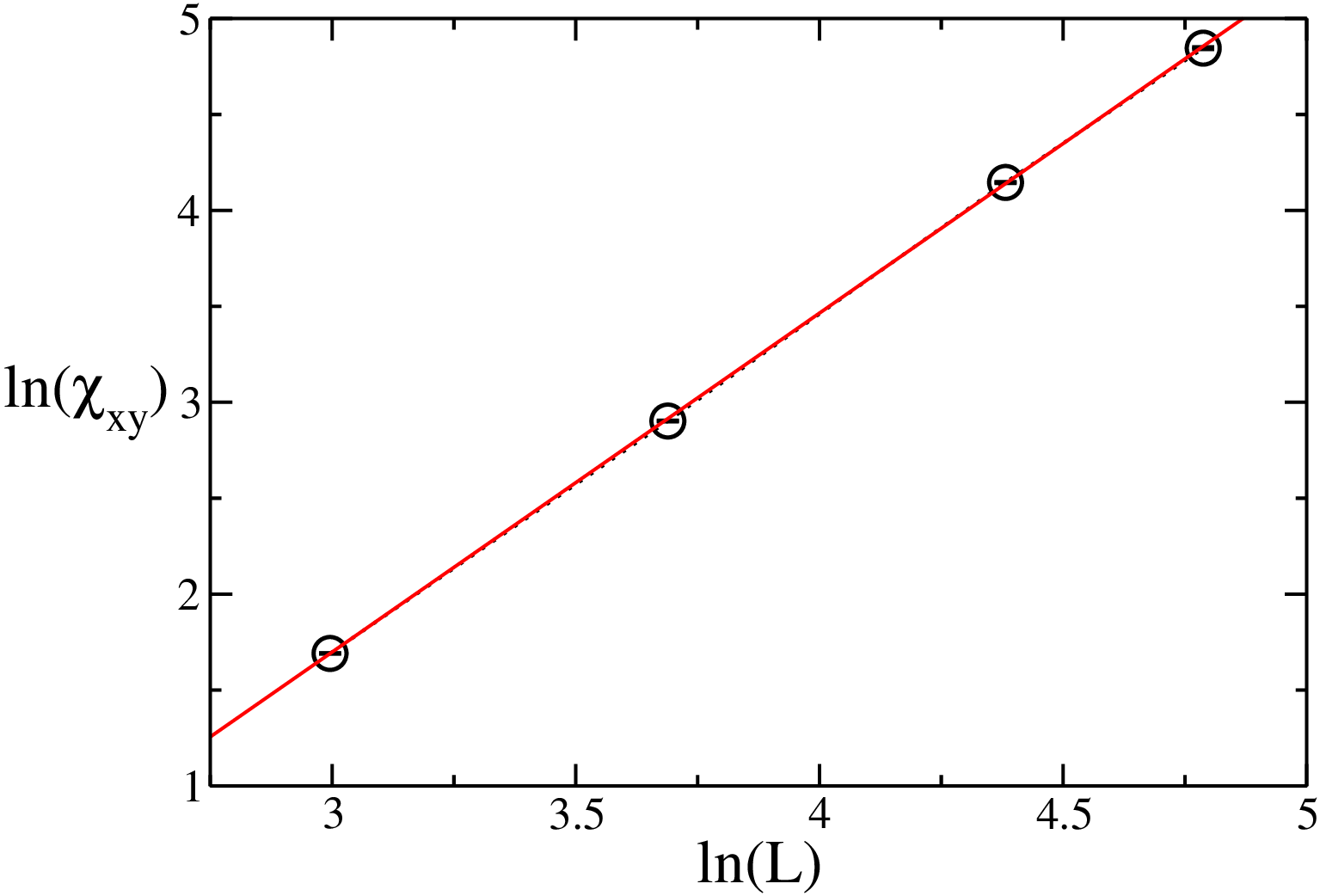}
\caption{\label{fss_sus} (Color online) Log-log plot of the maxima of the planar
susceptibility as a function of the lattice size. The solid red line shows the
best linear fit of the data given the exponent $\gamma /\nu=1.763(1)$.
The error bars are shown inside the symbols.}
\end{center}
\end{figure}

In figure~\ref{fss_sus} we show a $log-log$
plot of the maxima of the susceptibility as a function of the lattice
size for $L = 20, 40, 80$ and $120$. The data are very well adjusted
by a straight line with slope $\gamma/ \nu = 1.763(1)$ exhibiting
a power law behavior. This value of the exponent $\gamma /\nu$ is quite near
the expected one for a transition in the Ising universality class (1.75). 
Considering the Ising universality class we were able to determine the critical
temperature by using the location of the maxima of the specific heat and
susceptibility and the crossing point of the Binder's cumulant, which gives
$T_c^{Ising}=0.946(1)$. By using this value and plotting $\ln (M_{XY}) \times \ln(L)$
at $T=T_c$ we have found $\beta/\nu=0.163(6)$, which is quite different from
the expected value for the Ising universality class ($0.125$).
%
%========================= figure 2 ==============================
%
%==================================================================
%

The last result may indicate that the assumption of the Ising universality class may not be correct.
Indeed, by analyzing the moments of magnetization defined in Ref.~\onlinecite{mol2p}, we obtain
$1/\nu=0.82(2)$ and $T_c^{V_j}=0.943(1)$. This value
of the exponent $\nu$ contrasts with the expected for the Ising universality class, although 
the value for the critical temperature is approximately the same. Reanalyzing our previous 
estimates for the critical temperature obtained using the location of the specific heat
and maxima of susceptibilities using this new value of the exponent $\nu$ we obtain:
$T_c^{c_v}=0.945(1)$ and $T_c^{\chi}=0.943(1)$ (it is worthy to note that in the analysis
of the specific heat data the point corresponding to $L=20$ was disregarded in both cases).
Looking to the crossing point of the Binder's
cumulant we have found $T_c^{U_4}=0.944(2)$.
We have thus, as our new estimate for the mean critical temperature $T_c=0.944(1)$. 
Using this new value of the critical temperature we obtain $\beta/\nu=0.149(7)$ in the
analysis of the magnetization data.

\begin{figure}
%\vspace{0.5cm}
\begin{center}
\includegraphics[scale=0.25]{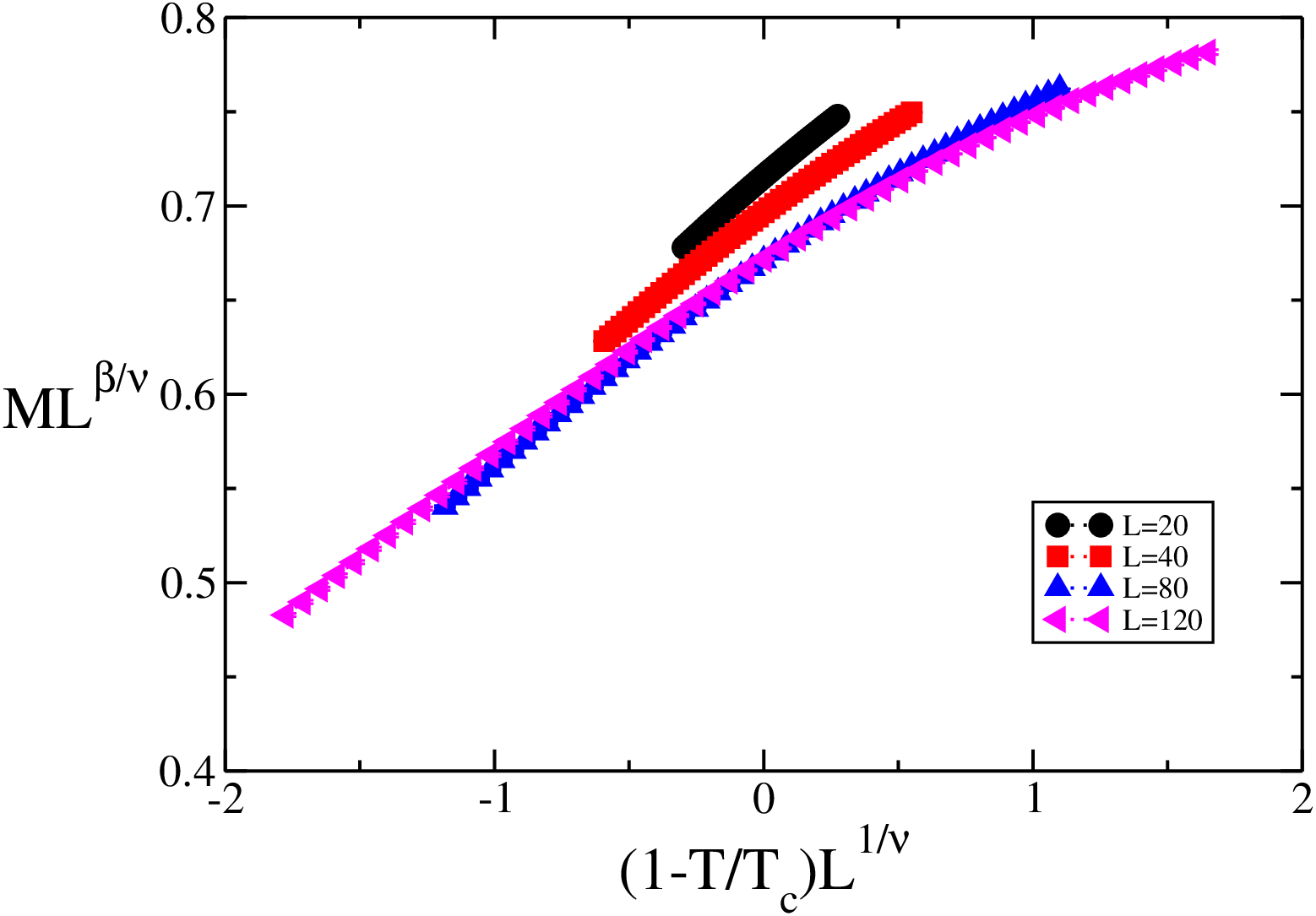} \\ \includegraphics[scale=0.25]{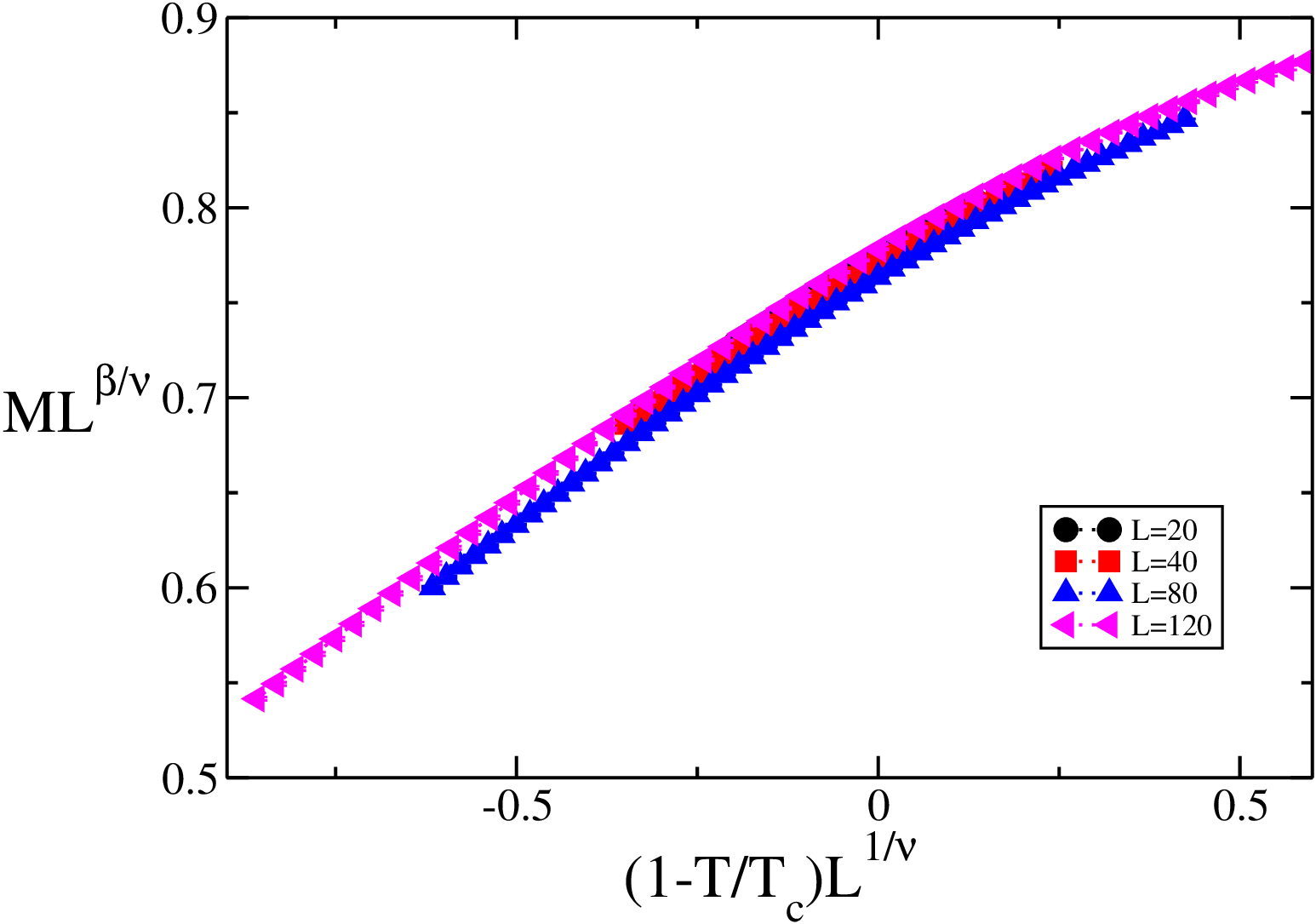}
\caption{\label{sc_mag} (Color online) Scaling plots of magnetization considering
the Ising-like behavior (top) and an order-disorder transition characterized
by the exponents shown in the last line of table~\ref{table} (bottom).
}
\end{center}
\end{figure}

To distinguish between these scenarios in figure~\ref{sc_mag} we show a
scaling plot of the magnetization obtained with the multiple histogram technique according to its finite size scaling function
($m \approx L^{\frac{\beta}{\nu}} \textbf{\textit{M}}\left( tL^{\frac{1}{\nu}} \right) $)
considering two possibilities: ($i$) the Ising-like behavior ($T_c^{Ising}=
0.946(1)$, $\nu=1$ and $\beta=0.125$) and ($ii$) an
order-disorder critical behavior with exponents $\nu=1.22(3)$ and $\beta=0.18(1)$ and critical
temperature $T_c=0.944(1)$. As can be seen, the scaling plot obtained
assuming the Ising universality class does not describe our data as good as the
results considering a new universality class. Besides, doing the same analysis with
susceptibility and Binder's cumulant no significant deviations were observed between
these two possibilities. Indeed, the values obtained in this study
are in good agreement with those obtained for the same model in a bilayer system\cite{mol2p}
and for the dipolar Planar Rotator model\cite{molxy}. To clarify, in table~\ref{table}
we show the exponents for the Ising model, the results obtained by Maier and Schwabl for the
dPR model, the results of Refs.~\onlinecite{mol2p,molxy} and the results of this work.

%
%========================= figure 3 ==============================
%
%==================================================================
%
%
\begin{table}\begin{center}
\begin{tabular}{lccccccc}
\hline \hline
Model          & $T_c$         & $\nu$ & $\gamma$      & $\beta$       & $\alpha$      \\
\hline
Ising           & 2.269         & 1     & 1.75          & 0.125         &  0 ($\ln$)    \\
dPR (Maier)     &               &       & 1             & 1/2           & -2            \\
AHd (bilayer)   & 0.890(4)      & 1.22(9)& 2.1(2)       & 0.18(5)       &  -0.55(15)    \\
dPR            & 1.201(1)      & 1.277(2)& 2.218(5)    & 0.2065(4)     &  -1.1(1)      \\
AHd (this work)            & 0.944(1)      & 1.22(3)& 2.15(5)    & 0.18(1)      &  -0.44(18)    \\
\hline \hline \end{tabular} \caption{ \label{table} {In this table we show the critical temperature
and exponents for the 2D Ising model\cite{onsager} (first line), the results of Maier and Schwabl\cite{maier}
for the dPR model, the results of MC calculations in the bilayer AHd model with a cut-off in the interactions
\cite{mol2p}, the results of MC calculations for the dPR model\cite{molxy} and the results of this work. }} \end{center}
\end{table}

\begin{figure}
%\vspace{0.5cm}
\begin{center}
\includegraphics[scale=0.25]{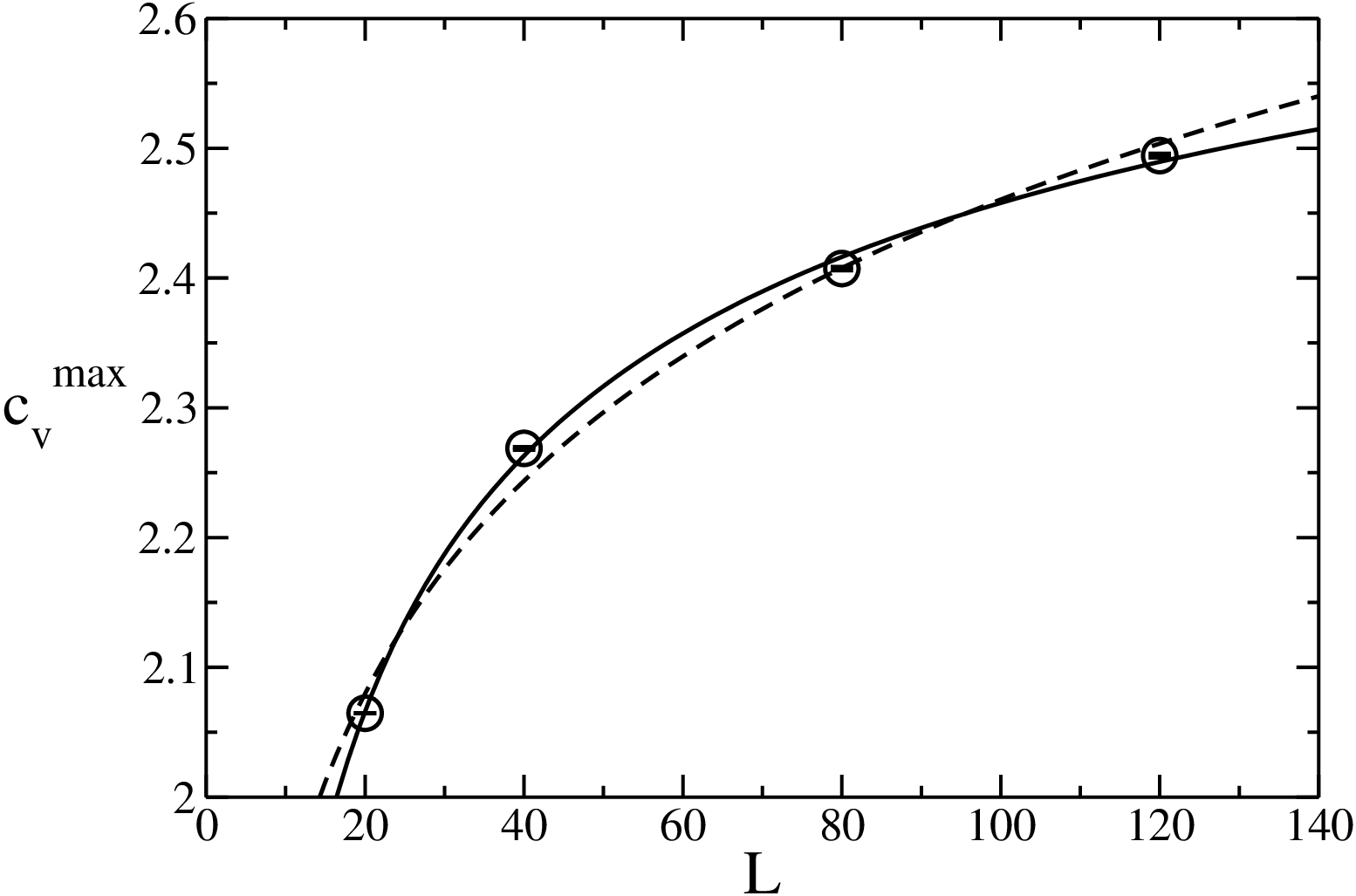}
\caption{\label{fss_cv} (Color online) Specific heat maxima as a function of the 
lattice size. The dashed line is the best non-linear fit considering an logarithmic 
divergence and the dashed line shows the best non-linear fit considering a non-divergent power law behavior.
}
\end{center}
\end{figure}

So far, everything corroborates to an order-disorder phase transition with non-conventional
critical exponents. However, the scale relations\cite{privman}
$\alpha + 2\beta + \gamma =2$ and $\nu d=2-\alpha$ are believe to be satisfied. Using the
values shown in table~\ref{table} and the first relation we should have $\alpha = -0.51(7)$ and using
the second relation $\alpha=-0.44(6)$ indicating the possibility that the specific heat
does not diverges. Indeed, to have an better agreement between the results of this work
and those of Refs.~\onlinecite{mol2p,molxy} the specific heat should be non-divergent.
As one knows, to distinguish between a logarithmic divergence or a slowly power law
divergence or even a non-divergent power law, much larger system sizes must be used. However, such analysis demands a prohibitive computer time. 
Nevertheless, a careful analysis of the data could give us a clue. In figure~\ref{fss_cv}
we show our data for the maxima of the specific heat as a function of the lattice size
adjusted by two different methods. The dashed line represents the best fit of a logarithmic divergence
($a\ln (L)+b$), the solid line is for a non-divergent power law behavior ($-a L^{-b}+c$). As can be clearly seen, the
non-divergent power law describes better the data. Indeed, the $\chi^2/dof$ values obtained
are $4.7\times 10^{-4}$ for the logarithmic divergence and $1.4\times 10^{-6}$ for the non-divergent
power law. The value obtained for the exponent $\alpha/\nu$ from the adjust is $-0.36(14)$, that agree quite well with the results of the present work and the simulations for the AHd model (See Table I).
%
%========================= figure 4 ==============================
%
%==================================================================
%

\section{\label{Conclusions}Conclusions}
In this work we have studied the phase transition in the
anisotropic Heisenberg model with dipolar interactions (AHd).
%Our main goal was to look for possible differences in the critical
%behavior of the planar to paramagnetic phase transition when full long-range
%dipolar interactions are considered or not. 
We have found that the use of the 
full long-range interaction 
leads to 
%different results, i.e., while the introduction of a cut-off radius at
%five lattice spacings lead to a BKT transition (Ref.~\onlinecite{marcella}),
%the use of the Ewald summation lead to 
an order disorder transition with unusual
exponents and a non divergent specific heat. Indeed, it would be interesting
to present a systematic study of the effects in the critical behavior of the
system with an increasing cut-off radius. Nevertheless, this study is
beyond the scope of this paper and will be addressed in a near future.

Since the analysis of the results presented in this paper is similar to those of Refs.~\onlinecite{mol2p,molxy}, we report the reader to those references for a more detailed discussion, specially to Ref.~\onlinecite{molxy}. Nevertheless, some points should be stressed.
In Ref.~\onlinecite{marcella} the authors have found that the planar to paramagnetic phase transition
in the AHd model belongs to the BKT universality class, which implies in the absence
of long-range order in the low temperature phase.
This result is consistent with the Mermin-Wagner theorem~\cite{mermin-wagner},
however, this theorem does not apply to systems with anisotropic long-range interactions as the AHd model.
Indeed, for such a system one should expect the existence of long-range order in the 
low temperature phase as shown by Maleev~\cite{maleev}. The main difference in the methodology
between Ref.~\onlinecite{marcella} and this work is that in the former a cut-off radius
were introduced in the evaluation of dipolar interactions while in the later the Ewald summation
was used. This is a clear indication that the inadvertent introduction of a cut-off
radius may hide the true critical behavior of the system. However, even with
the introduction of a cut-off radius, the results for a bilayer system~\cite{mol2p}
show the same critical behavior found in this work. This may indicate that
the anisotropic character of dipolar interactions is a key factor.
Indeed, the authors of Ref.~\onlinecite{fernandez} stated that ``Anisotropy has a deeper effect on the ordering of systems
of classical dipoles in 2D than the range of dipolar interactions'',
showing that this observation is not new in the literature. 

Although our results show that the unusual exponents shown in table~\ref{table} describes better
the data, specially for the magnetization, the transition may be in the Ising 
universality class as well, since corrections to scaling were not taken into account
and the lattice sizes used may not be large enough. As can be seen in
figure~\ref{sc_mag} the use of the Ising universality class exponents describes
well the data for the largest lattices studied. Nevertheless, it does not seems
to be a good choice simply disregard the data for the lattices with $L=20$ and $40$, 
leaving only two lattice sizes to be analyzed. Thus it is more prudent to not completely
rule out the possibility of this phase transition to belong to the Ising universality class.
On the other hand, our previous results for the same model in a bilayer system~\cite{mol2p} and the results
for the dPR model~\cite{molxy} were also well described by the same critical behavior found
here, such that we still believe that this phase transition is more likely to belongs
to a new universality class with unusual exponents.
Studies in much larger lattices could remove this ambiguity, nevertheless
the computational time needed for such a study turns it impracticable at the moment.

As a final remark we would like to stress that these results are much important
when the critical behavior of magnetic
systems with dipolar interactions is being considered. They show that the use
of a cut-off radius in the evaluation of dipolar interactions may lead to erroneous
results and that the anisotropic behavior is also much important. This study may be a guide for future works in what concerns the
introduction or not of a cut-off radius in the study of critical behavior of
magnetic systems with dipolar interactions.

\acknowledgments This work was partially supported by CNPq, FAPEMIG and PRPq/UFMG (Brazilian Agencies).

%
%------------------------------ Bibliography ------------------
%
\bibliography{AHD}
\end{document}